# Scalable and telecom single-erbium system with record-long room-temperature quantum coherence

**Authors:** Alex Kaloyeros, Natasha Tabassum, and Spyros Gallis*

**Affiliations:**

College of Nanotechnology, Science, and Engineering (CNSE), Department of Nanoscale Science and Engineering, University at Albany (UAlbany), NY 12222, USA

**Abstract:** Achieving long-lived (μs-scale) quantum coherence at room temperature in the telecom C-band remains a central unsolved challenge for solid-state quantum platforms, with coherence at this timescale previously demonstrated only under vacuum cryogenic conditions at temperatures over 600 times lower. To this end, we introduce a CMOS-compatible nanoengineering approach that deterministically controls the number of Er ions within a nanoscale volume and their electromagnetic environment using sub-5-nm silicon-based hollow nanopillars, which achieves the first deterministic spatial isolation of single erbium ions. This nanostructure-enabled isolation suppresses ion-ion interactions and local electromagnetic-field noise, enabling single-erbium-ion operation under ambient conditions. The resulting individually addressable single-ion devices exhibit record room-temperature optical coherence times of $T_2^* = \sim 32$ μs in the telecom C-band (limited by the coherence of the laser) and extending to > 100 μs based on photon echo measurements. We further present the first demonstration of background-free, upconversion-enabled single-photon emission from multilevel erbium qudits, providing high-contrast coherent optical readout. Beyond these performance gains, our work, through sub-5-nm nanoscale engineering, establishes isolated single-ion environments as a viable route to room-temperature operation for solid-state quantum systems. This approach has the potential to address the long-standing barrier of room-temperature decoherence and open a manufacturable, telecom-compatible pathway toward scalable quantum networks, distributed sensing, and cryogenic-free quantum photonic integrated circuits.

Advancing quantum information science (QIS) - including sensing, photonics, information processing, and communications - demands innovative materials engineering to overcome one of the central unsolved challenges of telecom quantum systems: the requirement of cryogenic operation due to rapid decoherence at elevated temperatures.[1-5] Thus, an interrelated challenge is to develop solid-state quantum systems that retain microsecond-scale[6, 7] quantum coherence at higher temperatures, ideally reaching room temperature (RT),[3, 4] while simultaneously supporting scalable, CMOS (complementary metal-oxide semiconductor)-compatible fabrication and telecom C-band operation. [1, 4] Thus far, no such system has been demonstrated. Tackling this long-lasting challenge is essential for transitioning from laboratory prototypes to practical, on-chip telecom quantum systems, including quantum photonic integrated circuits (qPICs),[8-10] and for meeting the size, weight, power, and cost (SWaP+C) constraints that deployable quantum technologies demand.[4] On-chip quantum systems, based on telecom quantum emitters and qubits (including, potentially, higher-dimensional qudits, with $d > 2$ energy levels) that operate in the low-loss telecom band (e.g., the C-band around 1540 nm) and are compatible with fiber networks,[6, 11-13] form the technological foundation of next-generation QIS applications, such as long-distance quantum networking, quantum information storage and processing, and distributed quantum sensing and computing.[13-15]

A range of solid-state quantum systems, such as semiconductor quantum dots, [16-18] defects,[19-21] and rare-earth-erbium ($Er^{3+}$) ions,[5-7, 22-27,] have enabled pioneering demonstrations of quantum functionality, such as state-of-the-art sensitivities,[28, 29] in measuring magnetic and electric fields in the near infrared range (1000 nm - 1300 nm), deterministic single-photon emission, spin-photon interfaces[25, 30] and microsecond-scale quantum coherence and high-fidelity optical control in the telecom C-band, [7, 24, 26, 27] albeit under controlled laboratory vacuum conditions at cryogenic temperatures. In these systems, individual $Er^{3+}$ ions are resolved spectrally rather than spatially, distinguished by optical detuning within a surrounding population of optically active ions. For example, many leading demonstrations approximate tens to hundreds of optically active $Er^{3+}$ ions within the interaction volume surrounding the probed ion.[7, 24, 26] Although spectral isolation enables selective optical addressing of a single ion, neighboring ions persist within its environment, leading to ion-ion coupling and local field fluctuations, and these decoherence pathways are commonly mitigated through cryogenic operation (<0.5 K) and applied magnetic fields. More generally, rare-earth ion coherence is limited by magnetic dipole-dipole coupling to neighboring electronic and nuclear spins, spectral diffusion arising from rare-earth-ion interactions, and local electric- and strain-field fluctuations associated with defects and structural disorder.[31, 32] Spatial isolation and nanostructuring offer a different approach by reducing the number of interacting ions and surrounding electromagnetic noise within the immediate environment of the emitter (**Supplementary Figure S1**). To our knowledge, spatial isolation of a single $Er^{3+}$ ion in a solid-state system has not been demonstrated at any temperature and constitutes a central enabler of the results reported here. Overall, scaling these systems beyond the lab remains elusive due to stringent operating conditions, a lack of native telecom C-band operation, and/or limited compatibility with large-scale fabrication, constraints that limit their suitability for practical, low SWaP+C telecom quantum technologies. **Supplementary Data Table S1** highlights the lack of a unified solution by summarizing these quantum systems (e.g., telecom-band operation, CMOS-compatible fabrication, and state-of-the-art coherence creation and control above cryogenic temperatures), motivating our work.

To this end, we introduce a nanoengineering approach that deterministically controls the number of Er ions within a nanoscale volume and their electromagnetic environment. By reducing the interacting-ion population from the ensemble regime to the single-ion limit, optical coherence increases by more than three orders of magnitude, culminating in long-lived room-temperature telecom-band optical coherence from spatially-isolated single $Er^{3+}$ ions. On this, we demonstrate spatial isolation of a single $Er^{3+}$ ion, enabling optical coherence times exceeding 100 μs in the telecom C-band at room temperature, realized through a foundry-compatible nanofabrication platform paired with materials engineering. This provides a route to a new class of CMOS-scalable solid-state quantum systems: arrays of spatially-isolated single-$Er^{3+}$-ion qudits embedded in silicon-based (e.g., silicon carbide (SiC) and $SiC_xO_y$) hollow nanopillars (HNPs). This is achieved through nanostructure geometries with ultrathin critical dimensions (≤5 nm), defined by conformal deposition rather than lithographic resolution, which provide geometry-driven confinement of individual $Er^{3+}$ ions within nanoscale volumes, thereby reducing neighboring-ion interactions and limiting spectral diffusion associated with local charge fluctuations and defects. SiC and $SiC_xO_y$, as host matrices for $Er^{3+}$ offer unique fundamental and silicon-technology advantages.[33-35] For example, amorphous-SiC-on-insulator (a-SiCOI) [36] offers CMOS compatibility, a wide bandgap, a high refractive index ($n \approx 2.6$ at 1550 nm), and strong electro-optic properties.[36, 37] Within our HNP devices, we observe a three-order of magnitude increase in the effective excitation cross-section ($\sigma$) for $Er^{3+}$-ions, enabling their isolation and coherent optical control under ambient conditions. These capabilities that have previously required nanophotonic cavities, cryogenic temperatures, and spectral isolation of Er ions.[5, 7, 22, 23] By systematically reducing the number of probed ions from ensembles (i.e., 60, 20, and 9 Er ions) to single ions, we show that coherence times increase substantially, by more than three orders, Rabi decay times lengthen, and Rabi contrast improves. Our single-ion devices characterization comprises photon-statistics measurements establishing spatial isolation on both the telecom transition and the visible upconversion, high-contrast Rabi oscillations (>96%) with no resolvable decay over the investigated pulse lengths, Ramsey and photon-echo sequences, and excited-state lifetimes determined independently by time-correlated single-photon counting (TCSPC), pulsed-$g^2$ measurements, and $\pi$ and $\pi/2$ pulse sequences.

Our quantum devices exhibit room-temperature performance ($T_2^* = 32$ μs) that is comparable to what has been previously reported in current technologies under cryogenic temperatures. [6, 7, 24, 26, 27, 38] These results demonstrate that long-lived coherent control is possible under practical, ambient conditions. Furthermore, we exploit the multi-energy-level structure of these spatially isolated single-Er qudits ($d$ = 5-level system) to realize background-free, single-photon emission via an upconversion-mediated readout pathway. The upconversion-mediated approach enables fast, high-contrast coherent optical readout in the visible (e.g., 520 nm) without the use of an optical cavity. Together, these results demonstrate a telecom- and CMOS-compatible engineering strategy, combining materials engineering and nanoengineering, for maintaining quantum coherence under ambient

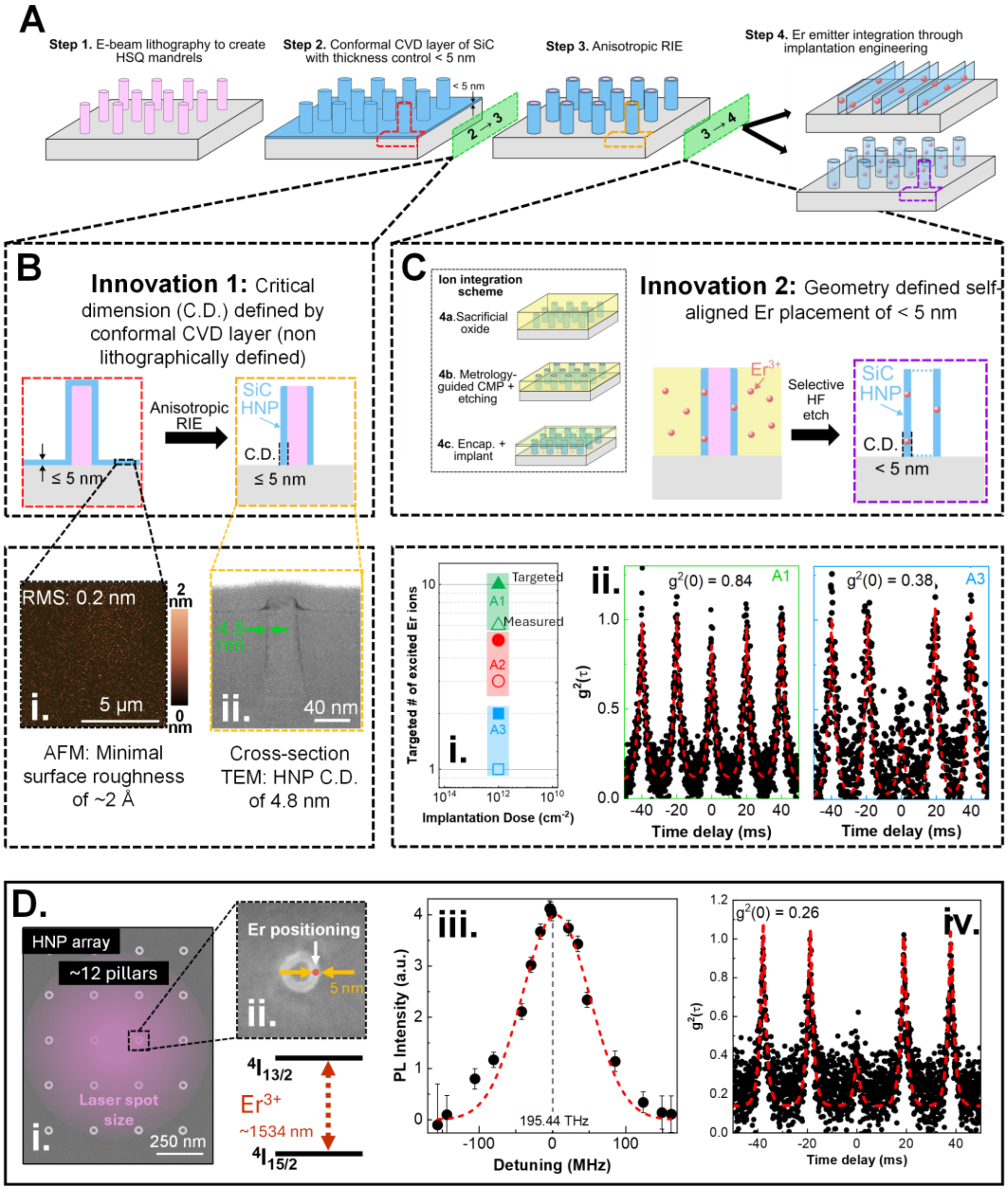

**Figure 1: Key-enabling nanofabrication for deterministic integration of single-Er-ions in Si-based hollow nanopillar (HNP) arrays. A.** Nanofabrication sequence and innovations for Er in engineered Si-based ($SiC_xO_y$) nanostructures, including HNP and NW arrays. **B.** *Innovation 1*: Highly conformal CVD enables C.D. control below lithographic resolution limits (≤5 nm), defined by deposition layer thickness rather than lithography. **i.** AFM image of Si-based films showing RMS roughness of ~2 Å. **ii.** Cross sectional TEM of an HNP with a C.D. (sidewall) of ~4.8 nm. **C.** *Innovation 2*: Geometry-driven self-aligned ion placement, where the nanostructure sidewall serves as an implantation aperture, confining ions with <5 nm accuracy. **i.** For HNP arrays, four designs target 10, 6, 2, and 1 optically active $Er^{3+}$-ions within a 4×3 array. **ii.** Representative $g^2(\tau)$ measurements, with $g^2(0)$ values ranging from 0.84 (multi-ion, A1) to 0.38 (A3), with fits shown as red dashed lines to Equation 4 (Methods). **D. i.** Top-down SEM image of a representative 3×4 HNP array. and **ii.** a single HNP highlighting Er confinement within the ~5 nm sidewall. **iii.** Room-temperature PLE spectrum of a single-$Er^{3+}$ in an A3 3×4 HNP array exhibiting the distinct telecom C-band optical $Z_1$ - $Y_1$ transition ~195.44 THz (~1533.9 nm) with a linewidth of ~67 MHz, limited by the tunable laser. Gaussian fit used to extract the FWHM is shown as red dashed line. **iv.** Corresponding $g^2$ correlation measurement showing a $g^2(0) = 0.26$, confirming single photon emission from a spatially resolved single $Er^{3+}$ ion. Red dashed line shows fitting to Equation 4.

conditions, at temperatures more than 600-times higher than those required by current cryogenic telecom quantum platforms. This work could facilitate advanced telecom QIS technologies that could potentially extend the range of practical applications in quantum sensing and imaging for biomedical applications, and in quantum sensing and reference systems (clock synchronization, distributed computation)[4] for a future Quantum Internet of Things.[13, 39]

Figure 1 outlines the fabrication strategy and key-enabling innovations underlying the Er-doped Si-based HNP array platform. Building on our previously reported approach for Er-doped SiC nanowire (NW) structures[40], this approach advances beyond ensemble implantation to create and control single $Er^{3+}$ ions. Innovation 1 (**Figure 1B**) demonstrates the non-lithographic route for defining Si-based HNPs with ≤5 nm critical dimension (C.D.), achieved through a conformal chemical vapor deposition (CVD) process[41] with accurately-controlled growth rates of ~0.4 Å/s. A sacrificial mandrel patterned by electron-beam lithography (pink in **Figure 1A**) serves only as a geometric scaffold, while the final nanostructure dimensions are set by the thickness of a conformally deposited ≤5-nm-thick Si-based layer (amorphous SiC (a-SiC) or silicon oxycarbide layer (a-$SiC_xO_y$)),[40] deposited on the mandrel, which defines the nanostructure dimensions independent of lithographic resolution. Subsequent selective reactive-ion etching yields hollow stoichiometric a-SiC (and a-$SiC_xO_y$) HNPs with smooth, uniform sidewalls, and sub-5-nm C.D.s, exhibiting minimal surface roughness (~2 Å RMS measured via atomic force microscopy). By decoupling C.D. control from lithography and avoiding the typical etching steps in top-down nanofabrication, our scheme circumvents conventional lithographic resolution limits and minimizes sidewall roughness, two inherent challenges in lithographic-pattern-transfer nanofabrication and thereby avoiding additional fabrication induced surface disorder.[42]

Deterministic integration of single ions with nanometer-scale spatial accuracy,[43] is a central requirement for scalable single-ion qubits, as on-demand spatial isolation can enable scalability and enhanced coherent control.[31, 44, 45] We achieve this by exploiting the geometry of the HNP whose C.D. (≤ 5 nm) enables geometry-defined ion placement with <5 nm accuracy (Innovation 2, **Figure 1C**). In this implantation scheme, the HNP sidewall functions as a lateral implantation aperture that constrains the final ion-placement volume. Following removal of a sacrificial oxide used during implantation, implanted Er ions are confined within the Si-based HNP nanostructures (hereafter referred to as Er:HNP or Er:NW). It is worth mentioning that this geometry-defined confinement strategy is broadly applicable to any class of ion or defect introduced via implantation. Control over ion placement is ensured through a metrology sequence that determines the etch-stopping point and the overlap between the implanted Er depth profile and the HNP height. To engineer the conditions necessary for single-ion isolation, we have developed a geometry- and dose-based design blueprint that quantitatively correlates implantation dose, post-implantation processing, and nanostructure volume. **Supplementary data Figure S3** summarizes the representative nanostructure arrays investigated in this study, designated as HNP1, along with the calculated number of excited $Er^{3+}$ ions for each configuration. As either the implantation dose or the effective nanostructure volume is reduced, the probability distribution of $Er^{3+}$ ions within the optical excitation volume transitions from a many-emitter ensemble to a regime dominated by single-ion probability. To mitigate the inherently stochastic nature of ion implantation and enhance experimental yield, the HNP1 arrays were designed so that multiple individual HNPs were illuminated simultaneously within the telecom laser excitation beam. Based on the calculated beam size at $\lambda = 1534$ nm and the NA of the objective lens = 0.9, ~12 HNPs fall within the FWHM diameter, considering $(0.51\lambda)/NA$, or ~26 considering the $1/e^2$ diameter $(0.82\lambda)/NA$, see **Figure 1D i**). This geometry increases the probability that at least one illuminated HNP contains a detectable single $Er^{3+}$ ion. Each HNP1 array comprises a $10^3 \times 10^3$ lattice of HNPs distributed over a 250 × 250 μm² area with a 250 nm pitch. Three representative arrays (A1, A2, A3), implanted with progressively reduced targeted $Er^{3+}$ numbers ($N_{\#}$) ranging from ~10 ions (A1) to ~2 ions (A3) per excitation volume, were used to systematically investigate the evolution of quantum emission as the number of addressed ions was reduced (**Supplementary data Figure S3** for more details). RT pulsed second-order photon-correlation ($g^2(\tau)$) measurements performed under resonant telecom excitation on representative sites from each array revealed antibunching, confirming quantum emission. Fitting the correlation functions using the standard pulsed-$g^2(\tau)$ model (Equation 4 in **Methods**)[46] yield $g^2(0)$ values consistent with the targeted $N_{\#}$ (where $g^2(0) = (N_{\#} - 1)/N_{\#}$). For example, a representative site on A1 exhibited $g^2(0) = 0.84 \pm 0.02$, corresponding to ~6 emitters, whereas a site in A3 exhibited $g^2(0) = 0.38 \pm 0.03$, consistent with emission from predominately ~1 - 2 emitters (**Figure 1C ii**). At another representative A3 site, pulsed-$g^2(\tau)$ measurements yielded $g^2(0) = 0.26 \pm 0.03$, demonstrating clear antibunching below $g^2(0) < 0.5$ and confirming single-photon emission from a spatially isolated $Er^{3+}$ ion. To our knowledge, this constitutes the first demonstration at room-temperature of spatially-resolved isolation of single-$Er^{3+}$-ions without relying on spectral isolation, typically achieved using a nanophotonic cavity. [5, 7, 22, 23, 47] Subsequent RT resonant-telecom photoluminescence-excitation (PLE) spectroscopy revealed an optical transition centered at ~1533.9 nm within the telecom C-band, with a measured linewidth of ~67 MHz (**Figure 1D iii**), limited by the broadening of the tunable telecom C-band laser. [48]

To determine optimal implantation parameters, we employed stopping-and-range-of-ions-in-matter (SRIM) simulations to model the depth distribution of implanted Er ions and accompanying profile of implantation-induced lattice damage. This SRIM-based framework for minimizing implantation-induced defects has been experimentally validated in our previous report on Er-implanted thin-film lithium niobate (Er:LN).[49] These simulations provide a quantitative relationship between Er implantation depth and the local density of implantation-induced vacancies within the volume surrounding each Er ion. By comparing the simulated vacancy density across implantation energies and oxide encapsulation thicknesses, we select conditions that place Er at the desired depth while reducing implantation-induced damage. This control is complemented by a materials-engineering strategy[50-52] aimed at improving the optical addressability of a few and single $Er^{3+}$ ions at room temperature. Specifically, the $Er^{3+}$ excitation cross-section, $\sigma$, is 2-3 orders of magnitude larger than typical values of ensemble rare-earth ions in bulk hosts. [40] **Figure S1** in the Supplementary illustrates the impact of this enhanced excitation cross-section by comparing resonant telecom power-dependent PL (PDPL) measurements from a representative Er:HNP array containing ~5 $Er^{3+}$ ions with those from a pristine LN reference sample containing a comparable number of ions.

We use nanofabrication-enabled controlled placement to systematically vary the number of $Er^{3+}$ ions within the probe volume, via nanostructure geometry and implantation design, and directly track the resulting evolution of optical coherence and Rabi dynamics (**Figure 2**). We determine the number of excited Er ions using the same approach discussed above via $g^2$ measurements and PDPL measurements (**Supplementary Information**). We have also performed upconversion-PL measurements to independently estimate the number of $Er^{3+}$ ions at each site from the observed count rates (see Methods, also further discussed in more detail in **Figure 4**). Sites targeting approximately 60 ± 17, 20 ± 6, 9 ± 1, and 1 $Er^{3+}$ ions exhibited upconversion PL count rates of ~38,100, ~19,450, ~10,250, and ~750 counts $s^{-1}$, respectively (**see Supplementary Figure S13**), matching the expected difference in the number of Er ions. We measure Rabi oscillations using telecom-resonant pulsed excitation while varying the excitation pulse width $t_p$ at a fixed photon flux ($\varphi = 5\times10^{19}$ $cm^{-2}\cdot s^{-1}$), with the time-integrated PL intensity recorded as a function of $t_p$ and fit[24] using

$$I(t_p) = A[1 - \exp\left(-\frac{t_p}{T_{2,Rabi}}\right)\cos(2\pi f_{Rabi} t_p)] \ , \quad [1]$$

where $t_p$ is the pulse width, $f_{rabi}$ is the Rabi frequency, $T_{2,Rabi}$ is the Rabi decay time, and $A$ is a fitting parameter. No oscillations are resolvable at ~60 ions, indicating an extremely short $T_{2,Rabi}$ decay (dephasing) time (blue dashed line denotes fit in the figure using Equation 1) below the detection limit of this measurement (~50 ns); at ~20 ions, oscillations exhibit a decay time of $T_{2,Rabi}$ = 2.1 ± 0.2 µs with a contrast of ~25% and $f_{rabi}$ = 727 kHz; at ~9 ions, this increases to $T_{2,Rabi}$ = 4.8 ± 1.3 µs with a contrast of ~48% and $f_{rabi}$ = 696 kHz; and at the single-ion limit, no decay is measurable over the investigated pulse lengths with a contrast of >96% and $f_{rabi}$ = 694 kHz (**Figure 3**). A corresponding increase is observed in the optical coherence time as the number of excited $Er^{3+}$ ions is reduced from approximately 60, to 20, 9, and ultimately one. The measured coherence increases from below the experimental detection limit to ~32 µs, representing at minimum a $10^3$-fold increase. This trend is reproduced using both two-pulse ($T_{2,PE}$) and three-pulse photon-echo ($T_2$) measurements using the π/2 (360 ns) and π (720 ns) pulses determined from the Rabi oscillations (**Methods and Supplementary Figures S11 and S12**). The two-pulse photon-echo sequence is well suited for measuring long optical coherence times because it is less sensitive to laser phase fluctuations.[22] For two-pulse photon-echo measurements, the integrated echo intensities (seen as a burst of light when the dipole moment has rephased) were fit [26, 27] to

$$I = I_0 e^{-2\left(\frac{\tau_{free}}{T_{2,PE}}\right)^x} \quad [2]$$

where $\tau_{free}$ is the free evolution time and x = 1 due to the observed single-exponential decay of the echo intensity.[53] The time-integrated decay of the intensity from the three-pulse photon echo measurements is fit [54] to

$$I = I_0 e^{-\frac{\tau_{free}}{T_2}} . \quad [3]$$

From these fits (dashed lines in **Figure 2**), coherence values were determined to be $T_2$ = 87 ± 11 µs and $T_{2,PE}$ = 79 ± 10 µs at 9 ions, $T_2$ = 32 ± 4 µs and $T_{2,PE}$ = 38 ± 3 µs at 20 ions, with no observable coherence at 60 ions (**Figure 2B**). In the two-pulse photon-echo measurements, dephased optical dipoles are coherently rephased to generate a temporally localized echo, providing an experimental signature of preserved optical coherence in the $Er^{3+}$ system.[55] The consistent trends observed in the photon-echo and Rabi measurements suggest that reducing the number of interacting $Er^{3+}$ ions toward a single isolated ion suppresses room-temperature decoherence. Sub-5-nm confinement within the HNP geometry provides nanoscale control of the local ion environment, such as ion-ion interactions and defect-, strain-, and charge-induced dephasing, enabling long optical coherence times at room temperature.

To directly probe single-ion quantum behavior, multiple isolated emission sites in a HNP array A4 (**Supplementary Figure S3 C**), which was engineered to target single isolated Er ions ($N_\#$ = 1), were fabricated and identified via upconversion-mediated confocal PL mapping. One representative site is shown in **Figure 3A,** with several others shown in **Supplementary data Figure S5**. Pulsed-$g^2$ measurements at this site, labeled Q1, under telecom-resonant excitation show pronounced antibunching with $g^2(0)$ = 0.09 ± 0.02, confirming single-photon emission from targeted, spatially isolated single $Er^{3+}$ ion (**Figure 3B**). Telecom time-resolved PL (TRPL) measurements by TCSPC reveal a biexponential decay with components $\tau_1$ = 0.3 ms and $\tau_2$ = 1.3 ms, yielding an average excited-state lifetime $T_1$ of 1.2 ms. An independent measurement of $T_1$ performed using a π/2 excitation pulse and monitoring the integrated photon counts as a function of delay after excitation yielded a lifetime of 1.3 ms (**Supplementary Figure S6**). Telecom PDPL measurements further yield an effective excitation cross-section of $\sigma = \sim 5 \times 10^{-17}$ $cm^2$ (with an onset saturation photon flux $\varphi_{sat} \simeq 10^{21}$ $cm^{-2}$ $s^{-1}$ (**Supplementary data Figure S6**) consistent with the ensemble ($N_{\#}$= 5 ions) Er:HNP structures (Supplementary data **Figure S1**). Under telecom-resonant pulsed excitation, coherent Rabi oscillations are observed from single $Er^{3+}$ ions and fit with the Equation 1 (**Figure 3C**). From the fit (red dashed line in the figure), a Rabi frequency, $f_{Rabi}$, of 694 kHz was extracted. Correspondingly, Ramsey measurements were performed at room temperature under telecom-resonant excitation (**Figure 3D**). The Ramsey sequence consists of two optical π/2 pulses, each with a width of 360 ns, determined from the Rabi studies, separated by a variable free-evolution interval $\tau_{free}$ (**Supplementary data Figure S7**). The Ramsey signal decays exponentially with increasing $\tau_{free}$, with the decay time of this intensity corresponding to characteristic $T_2^*$ coherence time. Fitting the exponential envelope to $e^{\frac{-\tau_{free}}{T_2^*}}$ (red dashed line in the figure) yields an optical coherence time, $T_2^*$, of 32 µs ± 4 µs, nearly identical to the estimated coherence time of the PPCL550 excitation laser (~32 µs) as determined from frequency-noise power spectral density (PSD) measurements, corresponding to an intrinsic linewidth of ~10 kHz. This suggests that the measured coherence time is limited by laser phase noise, and a narrower-linewidth excitation laser is needed to reveal the full coherence of the Er ion. The single-exponential behavior indicates that decoherence of the single-$Er^{3+}$ ion in HNP devices under pulsed optical excitation is predominantly governed by pulse-excitation-induced effects and subsequent radiative relaxation dynamics.[55] To further investigate the optical coherence of individual $Er^{3+}$ ions, three-pulse photon-echo measurements were performed as a function of free-evolution time ($\tau_{free}$), following the approach used for $Er^{3+}$ ensembles in **Figure 2** , utilizing an initial π-pulse to start the ion in the excited state prior to the pulse sequence (**Figure 3E and Supplementary data Figure S7**). The final π/2 pulse is applied to project the resulting refocused single-$Er^{3+}$ onto the excited-state. Considering the limitation mentioned above related to the laser coherence, fitting the data using Eq. 3 (red dashed line in the figure) yields a $T_2$ of 172 ± 18 µs, extending the optical coherence to > 100 µs (see Methods and **Supplementary data)**. As mentioned above, the influence of excitation-laser phase noise remains a limitation in determining the full intrinsic coherence time of the $Er^{3+}$ ion. Additionally, this three-pulse photon echo measurement is also performed without the initial π-pulse, which results in a coherence state being projected into the ground state by the final π/2 pulse (**Figure 3F**). Fitting the data yields a similar $T_2$ of 159 ± 28 µs. The observed coherence times under ambient conditions demonstrates the suppression of dephasing pathways in our Er:HNP (e.g., effects of ion-ion and ion-defect interactions) and showcases that quantum coherence can be preserved even in the absence of cryogenic cooling. As mentioned above, the intrinsic optical linewidth of the single $Er^{3+}$ in SiC HNP is limited by the broadening of the tunable-telecom-laser. However, based on the observed optical coherence time of $T_2^*$ = 32 µs, the corresponding optical linewidth is projected to be on the order of a few kHz $\left(\Delta\nu_0 = \frac{1}{\pi T_2^*} = \sim 10 \text{ kHz}\right)$.

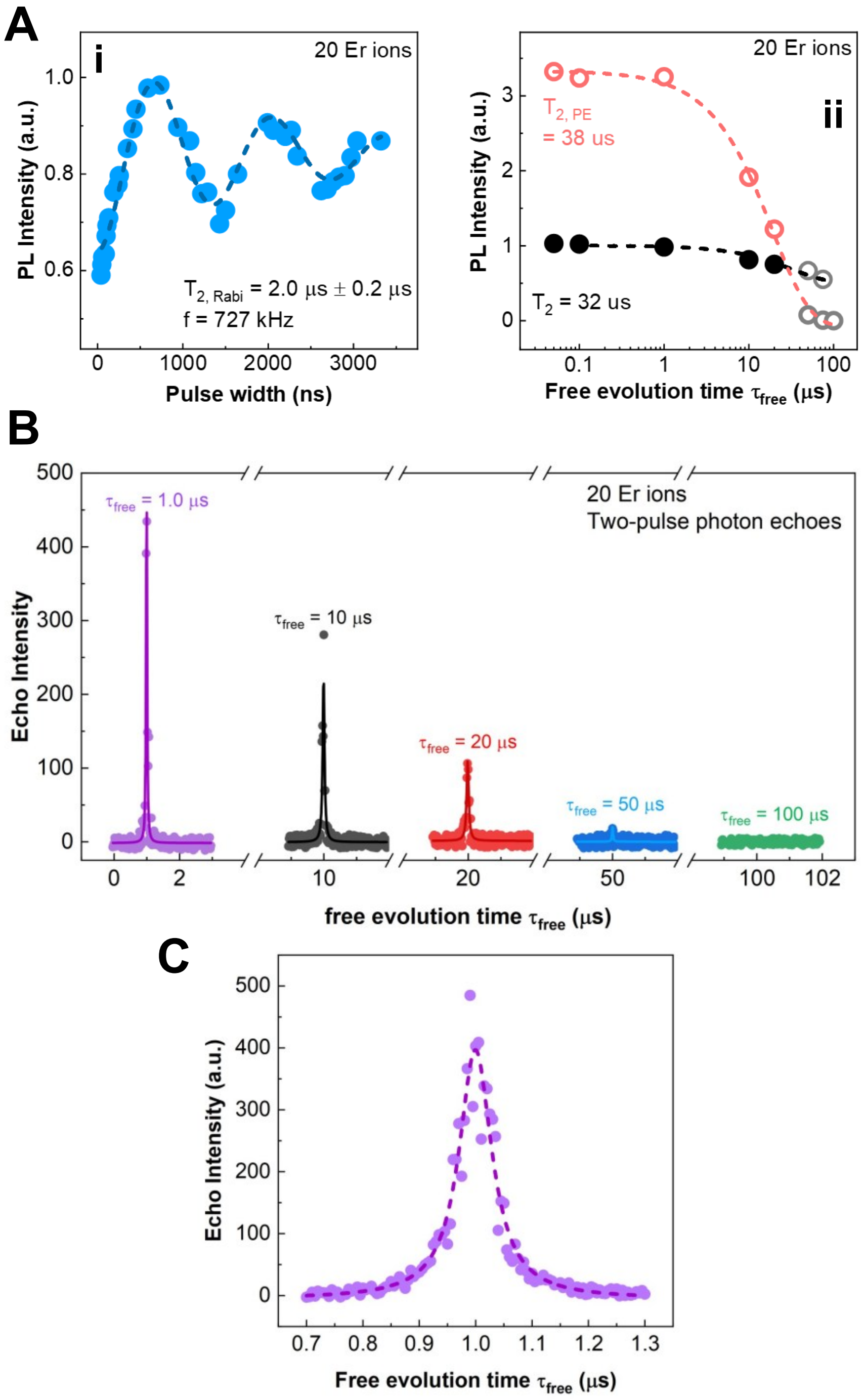

**Figure 2: Ion-number dependence of Rabi oscillations and optical coherence**. **A**. **i.** Rabi oscillation measurements (PL intensity as a function of pulse width) for spot containing 20 $Er^{3+}$ ions. Dashed line shows the fitting to Equation 1 used to extract $T_{2,\mathrm{Rabi}}$ and $f_{\mathrm{Rabi}}$. **ii.** Corresponding three-pulse ($T_2$, black) and two-pulse photon-echo ($T_{2,\mathrm{PE}}$, red) decay curves, plotted as PL intensity versus free evolution time ($\tau_{\mathrm{free}}$), for the same ion number. Dashed lines are fits to the exponential decays, yielding $T_2$ = 32 μs and $T_{2,\mathrm{PE}}$ = 38 μs. Gray open symbols indicate measurements at free-evolution times where the signal is likely to be influenced by the finite coherence time of the excitation laser. These points are shown for visual reference. **B.** Time-resolved telecom emission photon echoes from two-pulse photon echo from 20 Er ions for a $\tau_{\mathrm{free}}$ of 1.0 μs, 10, 20, 50, and 100 μs. **C.** Zoomed time-resolved telecom emission photon-echo from two-pulse photon-echo from 20 Er ions for a $\tau_{\mathrm{free}}$ of 1.0 μs. The dashed line represents a Lorentzian fit to the echo to extract integrated intensity.

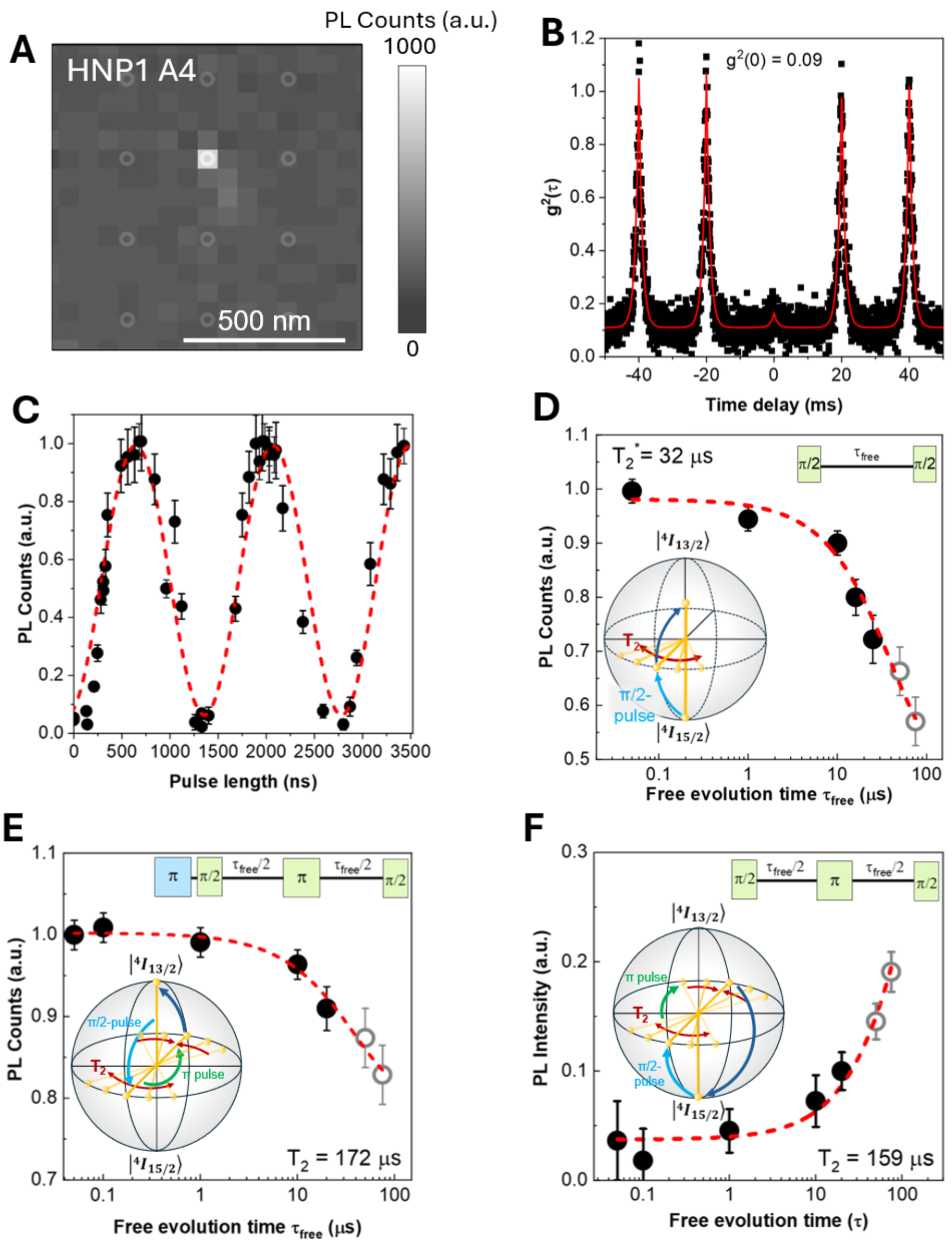

**Figure 3: Room temperature properties of single-Er SiC HNP devices in the telecom-C band. A.** SEM image of area representing single Er qubit Q1 on the HNP1 (A4) superimposed with upconversion-enabled PL camera image demonstrating spatial-isolation. **B.** $g^2(\tau)$ measurements for Q1 show pronounced antibunching under ~1534 nm excitation with $g^2(0)$ = 0.09, demonstrating telecom C-band single-photon emission. **C.** Rabi measurements on Q1 demonstrate coherent Rabi oscillations of the optical transition ($f_{Rabi}$= 694 kHz with contrast >96%). **D.** Ramsey spectroscopy on Q1 of the telecom Er transition: Integrated intensity decay of the $Er^{3+}$ emission after the Ramsey sequences. The inset schematically shows the decoherence process of the single-$Er^{3+}$ ion excited by a π/2 pulse; the dark-blue arrow indicates the second π/2 pulse of the Ramsey pair, which projects the single-$Er^{3+}$ state onto the excited state. The extracted decay time $T_2^*$ from a single-exponential fit is 32 μs. **E.** Corresponding three pulse photon-echo measurements varying the inter-pulse free-evolution time ($\tau_{free}/2$) with an initialization π pulse used to start the ion in the excited state (inset). A final π/2 pulse is applied to project the resulting refocused single-$Er^{3+}$ onto the excited-state. The photon-echo fit reveals $T_2$ = ~172 μs. **F.** Three-pulse photon echo measurement without the initialization π pulse. The photon-echo fit reveals $T_2$ = ~159 μs. Gray open symbols indicate measurements at free-evolution times where the signal is likely to be influenced by the finite coherence time of the excitation laser. These points are shown for visual reference. Red dashed lines denote fits.

To highlight the broader quantum functionality and technological promise of this platform, we demonstrate upconversion imaging of the $Er^{3+}$ emission, offering a pathway toward faster readout and sensing applications. We have implemented an upconversion-mediated pulse-sequence protocol to establish a background-free, high-efficiency optical readout pathway of the single-ion devices (Er:HNPs) and to access the multilevel structure of the qudit devices ($d$ = 5-level system) (**Figure 4**). In this scheme, resonant telecom-excitation at ~1533.9 nm using a sequence of $N$ $\pi$ pulses ($N$ = 2-3) populates higher-lying excited states, giving rise to visible and near-infrared upconversion emission at ~520 nm, ~660 nm, and ~980 nm (see Methods for more details). The corresponding energy-level diagram in **Figure 4A** illustrates potential excitation pathways, for example, of the $^2H_{11/2}$ manifold resulting in 520 nm upconverted emission excited using 3 sequential pulses. **Figure 4B** shows representative results from an isolated single-ion qudit (Q1). Upconverted PL at each wavelength is selectively collected using a corresponding bandpass filter and detected with a low-light single-photon camera (Andor DV437), enabling spatially resolved imaging of individual qudits with sub-diffraction resolution (<70 nm), limited by the camera pixel resolution (**Figure 4B i**). TRPL measurements of the upconverted emission yield lifetimes of 164 μs, 380 μs, and 712 μs for the 520 nm, 660 nm, and 980 nm channels, respectively from exponential fits (dashed lines in **Figure 4B ii**). These lifetimes are substantially shorter than the 1.2 ms lifetime of the $^4I_{13/2}$ state, consistent with radiative decay from higher-lying excited states involved in the upconversion process. PLE spectroscopy of the upconverted emission (shown here for the 520 nm emission) yields an optical linewidth of ~27 MHz determined from a Gaussian fitting (red dashed line in **Figure 4C**), limited by the excitation laser broadening.[48] This readout pathway supports coherent optical control, as evidenced by upconversion-mediated Rabi oscillations from individual $Er^{3+}$ ions (**Figure 4D i**). Pulsed-$g^2$ measurements of the upconverted emissions reveal, for the first time, enhanced high-$Q$ single-photon Er-emissions. Representative examples are presented in **Figure 4D ii** for the 520 nm upconversion states, exhibiting enhanced $g^2(0)$ values of 0.06 ± 0.01, reflecting reduced background contributions and improved signal-to-noise.

This work demonstrates the first CMOS-compatible platform of deterministic single-Er-ion qudits with coherent control at room temperature in the telecom C-band. By integrating single-$Er^{3+}$-ions into our sub-5 nm SiC HNPs devices, we address long-standing barriers to cryogenic operation, stochastic emitter placement, and non-scalable fabrication. We realize individually addressable single-Er-devices with record-long optical coherence times in the telecom C-band exceeding 100 μs at ambient conditions, a performance previously limited to vacuum conditions at temperatures over 600 times lower. Additionally, our quantum devices combine enhanced excitation efficiency with background-free, high-$Q$ upconversion-mediated single-photon emission, as demonstrated by Ramsey and photon-statistics measurements, establishing a pathway toward cryogenic-free qPICs for practical, scalable quantum technologies.

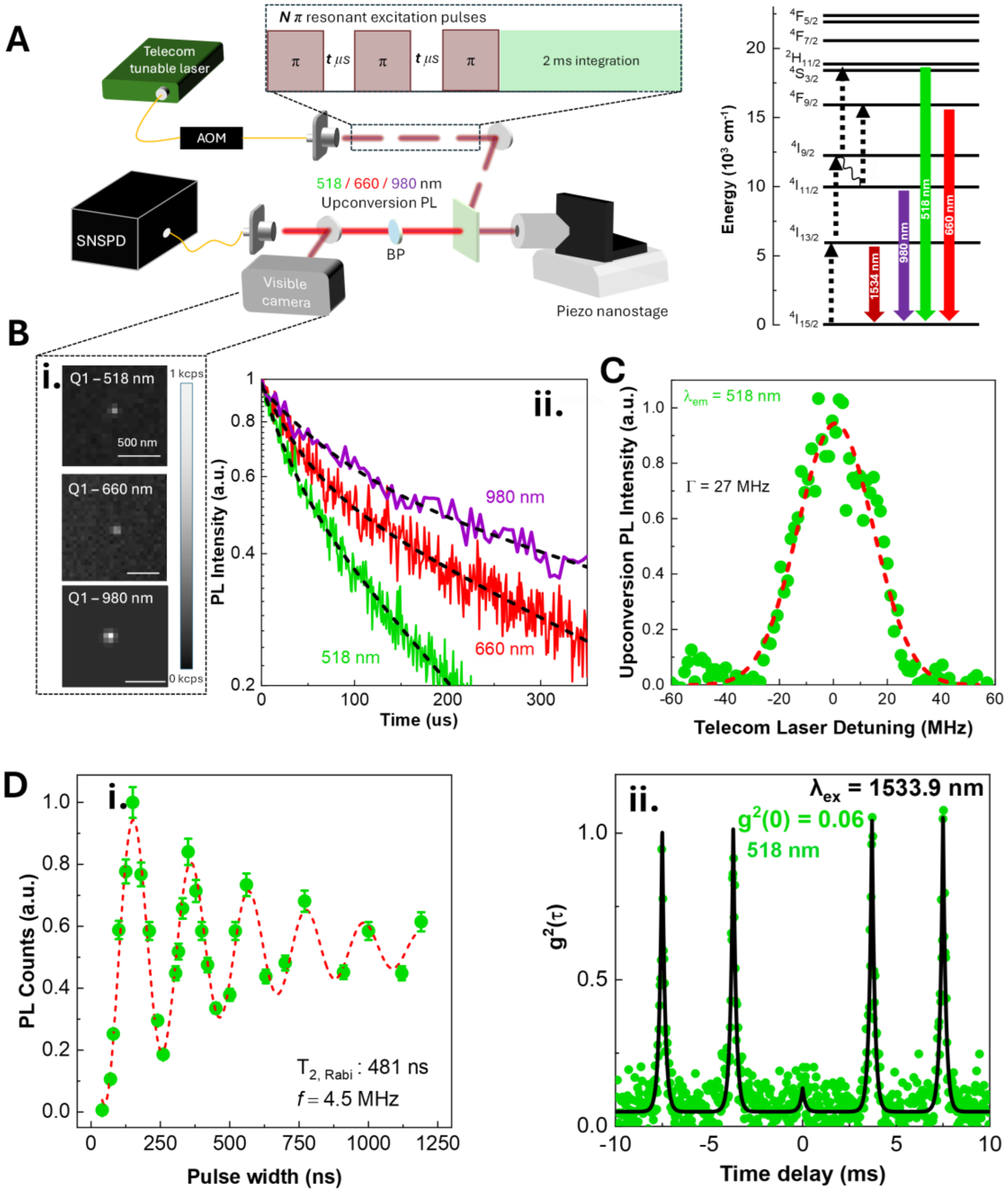

**Figure 4: Upconversion-mediated control of the visible and NIR states of single-$Er^{3+}$ qudits (5-level system). A**. Schematic of the measurement setup and pulsing sequence used for upconversion-mediated PL and microscopy from single-$Er^{3+}$ qudits. Two or three π telecom pulses resonant with the first excited energy level ($^4I_{13/2}$) with varied spacings are used to excite upconverted emission. Energy-level diagram for trivalent $Er^{3+}$ showing the telecom (~1534 nm) as well as its distinct visible (~520, $^4I_{15/2} \leftrightarrow {}^2H_{11/2}$, and ~660 nm, $^4I_{15/2} \leftrightarrow {}^4F_{9/2}$) and NIR (~980 nm, $^4I_{15/2} \leftrightarrow {}^4I_{11/2}$) optical transitions, involved in $Er^{3+}$ luminescence. **B**. Upconversion PL characterization of a single-$Er^{3+}$ qudit (Q1). **i.** Upconversion-mediated optical microscopy showing a single pixel (sub-diffraction) single-photon-emission spot corresponding to a single $Er^{3+}$-ion at three different wavelengths (520 nm, 660 nm, 980 nm). **ii.** Upconversion TRPL spectra showing emission at 520 nm, 660 nm, and 980 nm with lifetimes of 164 μs, 380 μs, and 712 μs, respectively. Dashed lines show exponential fits to extract lifetimes. **C.** Representative PLE with extracted optical linewidth of Q1 obtained via upconversion detection at 520 nm exhibits an optical linewidth of ~27 MHz at room temperature (laser-broadening limited). Red dashed line shows Gaussian fit. **D. i.** Upconversion PL at 520 nm excited using 3 sequential telecom pulses of varying width demonstrating Rabi oscillations under resonant excitation, revealing a π pulse of ~150 ns. Red dashed line shows the fitting to Equation 1. **ii.** Background-free and enhanced single-photon emission demonstrated for the single-$Er^{3+}$ qudit at 520 nm with $g^2(0) = 0.06$. Black line shows fitting to Equation 4.

**Acknowledgments:** The authors gratefully acknowledge Souryaya Dutta and Kazy Shariar for their assistance with materials growth, CMP, and etching during nanofabrication of the samples. This work was also supported by the College of Nanotechnology, Science, and Engineering (CNSE) Innovation Lab.

**Funding:** S.G. and A.K. were supported by the National Science Foundation under Grant No. 2138174-QuIC-TAQS. N.T. was supported by the National Science Foundation under Grant No. 1842350.

**Author contributions:** S.G. conceived and designed the experiments and methodology, including device nanofabrication, materials engineering, and quantum optics. A.K. and S.G. developed the experimental setup. A.K. fabricated the devices, performed the measurements, and analyzed the data. N.T. assisted with the growth, preparation, and measurements of 1st-generation nanostructures. S.G. assisted in device fabrication and quantum optics measurements. S.G. contributed to data analysis and advised on all efforts. A.K. and S.G. prepared the manuscript.

**Competing interests:** Three patent applications relating to this technology and innovations have been filed with the United States Patent & Trademark Office and are assigned to the Research Foundation for SUNY.

**Data, code, and materials availability:** The original contributions presented in this study are included in the article and supplementary material. Further inquiries can be directed to the corresponding author. Data reported in the supplemental are excluded in this preprint and will be made available upon publication of the peer-reviewed article or upon request.

## Methods

### Nanofabrication and metrology

**Supplementary data Figure S8** illustrates the nanofabrication process used to realize the ordered arrays of Er-doped silicon-based (e.g., silicon carbide (SiC) and $SiC_xO_y$) nanostructures (hollow nanopillars (HNPs) and nanowires (NWs)), referred to as Er:HNP or Er:NW. The process is compatible with CMOS manufacturing and enables independent control over the nanostructure geometry (diameter and critical dimension (C.D.) thickness), array periodicity, and controlled ion placement, potentially for any ion or defect compatible with the ion implantation process. While EBL is used here for rapid prototyping and device iteration, the device layouts and mandrel process are compatible with 300-mm wafer-scale lithographic manufacturing. For the structures presented in this work, the fabrication sequence begins with spin-coating a ~130 nm thick hydrogen silsesquioxane (HSQ) negative tone resist film onto a variety of substrates, including crystalline Si (e.g., Si (100) and Si (111)), thermally grown $SiO_2$-on-Si, and amorphous SiC on an insulator (a-SiCOI). Following a soft bake at 80°C for 4 minutes, electron beam lithography (EBL) is used to define high-aspect ratio (~4:1 height:width) cylindrical or wire-like HSQ mandrels, which serve as sacrificial templates for subsequent nanostructure formation. Patterned samples are developed in a tetramethylammonium hydroxide (TMAH) solution bath, yielding a periodic HSQ ribbon array (pink in **Figure S8**). An amorphous SiC (a-SiC) or oxygen co-doped a-SiC layer (a-$SiC_xO_y$, silicon oxycarbide) is subsequently deposited conformally by chemical vapor deposition (CVD) at 800°C (Step 2). The resulting film thickness is precisely controlled via growth time (e.g., ~4.8 ± 0.2 nm for the HNP arrays (as determined by ellipsometry, scanning electron microscopy (SEM), and transmission electron microscopy (TEM)). The film stoichiometry can be continuously tuned from stoichiometric a-SiC to stoichiometric a-$SiC_xO_{2(1-x)}$ ($SiC_{0.5}O_{1.0}$) by adjusting oxygen gas flow rate during growth, as detailed in previous reports.[50] This conformal deposition defines the sidewalls around the HSQ mandrels, which ultimately set the C.D. of the resulting HNPs. A fluorine-based anisotropic reactive ion etching process (RIE) is used to remove excess deposited material from the substrate surface and the tops of the mandrels while preserving the conformal sidewalls. This step converts the deposited film into vertically aligned HNP structures that adopt the underlying cylindrical or wire mandrel geometry. The nanostructure width (or C.D.) is therefore dictated by the deposition thickness, while the diameter and array periodicity are dictated by the HSQ mandrels. **Supplementary data Figure S8 B and C** present angled and cross-section SEMs of the Si-based HNPs, respectively.

To enable deterministic integration of $Er^{3+}$ ions, the fabrication process is extended beyond Step 3 following the RIE. An approximately 800 nm-thick sacrificial oxide layer is deposited by CVD to fully encapsulate the nanostructures and the remaining HSQ mandrels. The oxide surface is planarized using a multi-step chemical-mechanical polishing (CMP) process. A coordinated sequence of RIE steps and diluted hydrofluoric (DHF) etching, guided by a metrology protocol, is then used to recess the oxide to the precise height of the nanostructures. A second conformal oxide encapsulation layer is subsequently deposited on the planarized sample surface to control the ion-implantation range and, consequently, the number of ions incorporated into the nanostructures. In this study, Er ions are implanted at room temperature with an acceleration voltage of 100 kV and fluences ranging from $10^{12}$ to $10^{14}$ $cm^{-2}$. Post-implantation thermal annealing is performed at 900°C for 1 hour under argon and/or oxygen ambient to activate the $Er^{3+}$ ions and repair implantation-induced lattice damage. Following annealing, both the sacrificial oxide and the encapsulation oxide layers can be removed using a DHF etch, along with the remaining HSQ. This etch process exhibits high selectivity between the oxide layers and the SiC nanostructures, with an etch-rate contrast exceeding $10^3$ nm $s^{-1}$, leaving the Er:HNP or NW structurally intact, showing no measurable SiC etching under the conditions used in this work. **Supplementary data Figure S9** presents representative PL results for an ensemble Er:NW array.

**Calculations of system efficiency**

To estimate the collection efficiency of the 100× objective (NA = 0.9), we treat the emitter as an isotropic point source. For an objective in air ($n \approx 1$), the numerical aperture defines the maximum collection angle via $NA = n \sin(\theta)$, giving $\theta = \arcsin(0.9) = 64°$. The corresponding solid angle is $\Omega = 2\pi(1 - \cos\theta)$, so the fraction of photons collected relative to the full $4\pi$ emission is: $\eta = (1 - \cos\theta) / 2 = (1 - 0.44) / 2 = 0.28$. Finite difference time domain simulations of the vertical extraction efficiency $\eta_{ext}$ in the HNPs show that ~0.40 of the emitted power is extracted into the upper half-space. Within that upper half-space, the objective subtends a fraction $1 - \cos\theta = 0.56$, giving a net photonic collection efficiency: $\eta_{coll} = 0.40 \times 0.56 = 0.22$. This is multiplied by the measured transmission and detection efficiencies of each component of the microscope at ~1534 nm: objective (40%), beam splitter (99%), filter (97%), fiber coupling (87%), and SNSPDs (50 - 85%, depending on optimal vs anti-optimal polarization). The total system efficiency is therefore: $\eta_{sys} = 0.22 \times 0.40 \times 0.99 \times 0.97 \times 0.87 \times (0.5 - 0.85) = 0.037 - 0.062$, indicating that between ~3.7% - 6.2% of the photons emitted by the sample are ultimately detected.

**Experimental Setup**

The home-built confocal micrvo-photoluminescence (μ-PL) setup used in this work is shown schematically in **Supplementary data Figure S7**. Excitation is provided by a narrow-linewidth tunable laser in the telecom band (PurePhotonics PPCL550) or in the visible via an argon-ion laser (Spectra Physics Beamlok 2060) or diode lasers from Omicron (LuxX 457, 488), housed in a dedicated excitation module. The laser output passes through an appropriate band-pass filter set to further clean up the beam. The beam is then expanded to fill the back aperture of the objective, yielding a diffraction-limited spot on the sample. A Glan-type polarizer (Polarizer 1) cleans up the excitation polarization. The beam is reflected by a 99:1 mirror, with 1% of the laser power directed to the sample module and focused onto the sample by a Zeiss 100x, NA = 0.9, objective lens. The sample is mounted on a closed-loop piezoelectric nanostage (Physik Instrumente) with step resolution of < 2 nm. PL from the sample is collected by the same objective and directed back through the 99:1 transmission beam splitter into the detection module. In resonant telecom PL measurements, the tunable laser is pulsed, and the emission is collected within a time window after the end of the laser pulse. The SNSPD detector bias is shut off during the excitation pulse to prevent latching of the detector. For upconversion PL, a 520, 660, or 980 nm bandpass emission filter is used for the corresponding upconversion measurement. For spectrally resolved measurements, the light is coupled into a single-mode fiber, which sends the signal to a grating spectrometer connected to a pair of superconducting nanowire single-photon detectors (SNSPDs) (Quantum Opus) or a visible PMT. An Andor DV437 camera is used to image the PL. For second-order correlation measurements ($g^2(\tau)$), photoluminescence excitation (PLE), or coherence measurements, the emission is directly fiber-coupled to the pair of SNSPDs. These detectors provide high detection efficiency (>85%) and low dark count rates (<40 cps) in the telecom C-band and are interfaced to time-tagging electronics from Swabian Instruments (Time Tagger 20).

All telecom measurements were performed using resonant, pulsed optical excitation of the $Er^{3+}$ transition at 195.4 THz (telecom C-band) and a power of ~90 nW corresponding to a photon flux of ~$5 \times 10^{19}$ $cm^{-2}s^{-1}$, followed by detection of PL using SNSPDs. Upconversion measurements are done at an average photon flux of ~$2 \times 10^{22}$ $cm^{-2}s^{-1}$. PL detection consisted of a time-delayed fluorescence collection window after the excitation pulse is turned off. The linewidth of the tunable laser was broadened by its intrinsic frequency modulation response to approximately 100 MHz under a pulsing repetition rate of ~100 Hz for telecom C-band detection experiments and approximately 50 MHz under a pulsing repetition rate of ~500 Hz for upconversion-related measurements.[48]

For Rabi measurements, the system was driven with resonant optical pulses with variable duration $t_p$. Following each pulse, PL was collected during a fixed 3 ms integration window following a 500 ns delay from the end of the excitation pulse, with a dead time of 5 ms after the integration window. For upconversion measurements, 2-3 sequential resonant telecom pulses at ~1533.9 nm with 5 µs spacing between pulses are used to excite the Er ions, followed by a detection window of 2 ms with a dead time of 1 ms. The total photon counts were recorded as a function of pulse duration.

Ramsey spectroscopy was employed to probe optical coherence times. Each sequence consisted of two resonant $\pi/2$ pulses (360 ns duration), separated by a variable free evolution time $\tau_{\text{free}}$. PL was collected during a fixed 5 ms integration window following a 500 ns delay from the end of the excitation pulse. A 5 ms dead time was included between repetitions to allow full relaxation to the $^4I_{15/2}$ ground state.

At the frequencies relevant to the Ramsey and echo pulsing sequence used in this work, the Pure Photonics PPCL550 tunable C-band laser has a noise floor of ~1000 $Hz^2/Hz$, corresponding to a coherence time of ~80 - 100 µs, determined from frequency-modulation noise power spectral density (PSD) measurements referenced to a calibrated gas-cell absorption standard.[57, 58, 68] This decreases to ~32 μs based on the frequencies used in our coherence measurements.

The three-pulse photon-echo sequence was preceded by a resonant optical $\pi$-pulse used to initialize the system in the $^4I_{13/2}$ excited state. The sequence then consisted of a π/2 - π - π/2 sequence, where the π/2 and π pulses are separated by free evolution time $\tau_{\text{free}}/2$. All pulses were resonant with the $^4I_{15/2} \leftrightarrow {^4I_{13/2}}$ $Er^{3+}$ optical transition and used the same calibrated pulse areas as determined from the Rabi measurements. Following the final π/2 pulse, PL was collected during a fixed 5 ms integration window following a 500 ns delay from the end of the excitation pulse. A 5 ms dead time was included between successive sequences to ensure full relaxation to the $^4I_{15/2}$ ground state. The PL counts were extracted as a function of $\tau_{\text{free}}$ and fit to an exponential decay to determine the coherence time $T_2$. The final π/2 pulse is applied to project the resulting refocused single-$Er^{3+}$ onto the excited-state, in the absence of detuning and dephasing, or onto a coherent-superposition-state, as a result of decoherence. Additionally, this same three-pulse photon echo sequence was also performed without that initial optical $\pi$-pulse. In this case, the final π/2 pulse is applied to project the resulting refocused single-$Er^{3+}$ onto the ground-state. A two-pulse photon-echo sequence was also performed on spots containing ensembles of Er ions by applying a π/2 pulse followed, after a variable free evolution time, by a π pulse. The resulting echo was detected after a second equal free evolution time, and its intensity was recorded as a function of $\tau_{\text{free}}$ to determine the optical coherence time. The two-pulse photon-echo sequence is well suited for measuring long optical coherence times because it is less sensitive to laser phase fluctuations[22], variations in the optical phase of the refocusing π pulse alter the phase of the emitted photon echo but not its intensity, which is measured here.

Pulsed second-order correlation ($g^2$) measurements were performed with a resonant $\pi$ pulse (720 ns duration). PL was collected during either a fixed 3 ms or 10 ms integration window following a 500 ns delay from the end of the excitation pulse. For upconversion measurements, 2-3 sequential resonant telecom pulses (150 ns or 720 ns duration) at ~1533.9 nm with 5 or 200 μs spacing between pulses are used to excite the Er ions, followed by a detection window of 2 ms with a dead time of 2 ms. Temporal gating was implemented using the excitation pulse to define the start and stop of the detection window for both SNSPD channels. Coincidence events between the gated channels were used to construct the second-order correlation function $g^{(2)}(\tau)$.

Correlation functions were fit using the pulsed-$g^2(\tau)$ model[46]

$$g^2(\tau) = y_0 + \sum_i A_i exp\left(\frac{\tau - iT_{rep}}{\tau_{PL}}\right) \qquad [4]$$

where $y_0$ is a baseline offset, $A_i$ is the amplitude of the i-th peak, $iT_{rep}$ is the center of the i-th peak based on the laser pulse repetition rate, and $\tau_{PL}$ is the decay of the peak, which is related to the excited-state lifetime. We determine $g^2(0)$ as the ratio between the integrated area of the zero-time delay peak and the average integrated area of the side peaks. **Supplementary data Table S2** shows a comparison between the extracted excited-state lifetime $T_1$ and $\tau_{PL}$ values between the TRPL measurements and $g^2$ measurements, respectively.

**Table S2.** Comparison of excited-state lifetime between TRPL and $g^2$ measurements

| **Measurement/ Spot (emission wavelength)** | **$T_1$ (ms) from TRPL** | **$\tau_{PL}$ (ms) from $g^2$** |
|---|---|---|
| **Q1 (1534 nm)** | 1.2 | 1.2 |
| **Q1 (518 nm)** | 0.2 | 0.2 |
| **Q1 (980 nm)** | 0.7 | 0.6 |
| **Q4 (1534 nm)** | 1.2 | 1.4 |
| **A2 (1534 nm)** | 2.5 | 2.6 |
| **A3 (1534 nm)** | 1.6 | 1.5 |

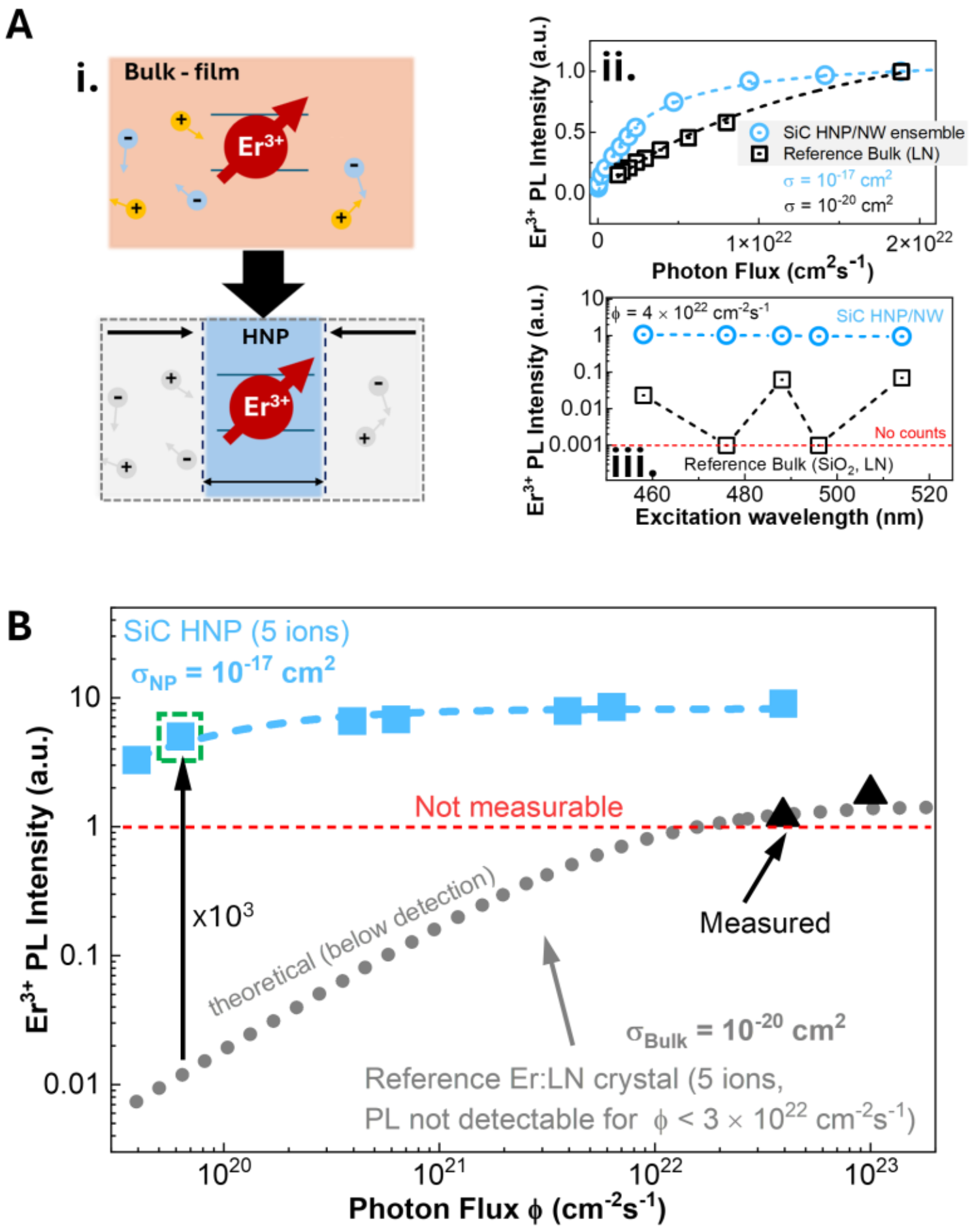

**Supplementary Figure S1: Enhanced excitation cross-section $\sigma$, for $Er^{3+}$ in SiC HNP arrays enabling room-temperature single-$Er^{3+}$-ion detection**. **(A) i.** Schematic illustrating how spatial isolation and nanostructuring of $Er^{3+}$ ions reduces coupling to defects, charge fluctuations, and phonon density of states compared to bulk or thin-film environments. **ii.** Telecom-PDPL studies and fits demonstrate that ensemble $Er^{3+}$ in the HNP/NW has $\sigma \approx 2 \times 10^{-17}$ cm$^2$, which is 2-3 orders of magnitude higher than typical values in bulk rare-earth hosts; for example, as seen in the Er-implanted LN thin films (reference). **iii.** Broadband excitation behavior of ensemble $Er^{3+}$ in HNP/NW compared to ensemble bulk reference (Er:LN and Er:$SiO_2$), as demonstrated by PLE at non-resonant (476 nm and 496 nm) excitations while monitoring the telecom emission. **(B)** Representative $Er^{3+}$-PL from Er:HNP array, containing ~5 $Er^{3+}$, exhibits an intensity ~3 orders of magnitude higher than the calculated intensity from the reference at the primary photon-flux ($\varphi$ ~$5\times10^{19}$ cm$^{-2}\cdot$s$^{-1}$) used in the experiments (telecom-resonant excitation ~1533.9 nm). In contrast, the PL in the reference, with an equivalent number of Er ions, remains essentially undetectable even at the highest excitation photon flux used, $\varphi > 9\times10^{22}$ cm$^{-2}\cdot$s$^{-1}$, a three-order higher value than the primary flux.